\documentclass{new_tlp}
\usepackage{mathptmx}
\usepackage{amsmath,amssymb}
\usepackage{latexsym}
\usepackage{stmaryrd}
\usepackage{graphicx}
\usepackage{hyperref}
\usepackage{listings}
\usepackage{verbatim}
\usepackage{xspace}
\usepackage{multicol}
\usepackage{color}
\usepackage{booktabs}
\usepackage{caption}
\usepackage{setspace}
\usepackage{cleveref}
\usepackage{subfig}
\usepackage{multirow}
\usepackage{soul}
\usepackage{boxedminipage}
\usepackage{wrapfig}

\newtheorem{example}{Example}

\newcommand\bcmdtab{\noindent\bgroup\tabcolsep=0pt%
  \begin{tabular}{@{}p{10pc}@{}p{20pc}@{}}}
\newcommand\ecmdtab{\end{tabular}\egroup}

\newcommand{\Symbols}{\Sigma}
\newcommand{\narval}{{\sf Narval}}%

\newcommand{\cR}{\mathcal{R}}
\newcommand{\cC}{\mathcal{C}}

\newcommand{\cT}{\mathcal{T}}
\title[Symbolic Analysis of Maude Theories with Narval]{Symbolic Analysis of Maude Theories with Narval\thanks{This work has been partially supported by the EU (FEDER) and the Spanish MCIU under grant RTI2018-094403-B-C32, by the Spanish Generalitat Valenciana under grants PROMETEO/2019/098 and APOSTD/2019/127, and by the US Air Force Office of Scientific Research under award number FA9550-17-1-0286.}
}

\author[M. Alpuente et al.]
         {MAR\'{I}A ALPUENTE, SANTIAGO ESCOBAR, JULIA SAPI\~{N}A\\
         VRAIN (Valencian Research Institute for Artificial Intelligence), Universitat Polit\`ecnica de Val\`encia\\
         \email{\{alpuente,sescobar,jsapina\}@upv.es}
         \and
         DEMIS BALLIS\\
         DMIF, University of Udine\\
         \email{demis.ballis@uniud.it}
}

\pagerange{\pageref{firstpage}--\pageref{lastpage}}

\begin{document}

\label{firstpage}

\maketitle

\begin{abstract}
Concurrent functional languages that are endowed with symbolic reasoning capabilities such as Maude offer a high-level, elegant, and efficient approach to programming and analyzing complex, highly nondeterministic software systems. Maude's symbolic capabilities  are based on equational unification and narrowing in rewrite theories, and provide Maude with advanced logic programming capabilities such as unification  modulo user-definable equational theories and symbolic reachability analysis in rewrite theories. Intricate computing problems may be effectively and naturally solved in Maude thanks to the synergy of  these recently developed symbolic capabilities and  classical Maude features, such as: (i) rich type structures with sorts (types), subsorts, and overloading; (ii) equational rewriting modulo various combinations of axioms such as associativity, commutativity, and identity; and (iii) classical reachability analysis in rewrite theories. However, the combination of all of these features may hinder the understanding of Maude symbolic computations for non-experienced developers. The purpose of this article is to describe how programming and analysis of Maude rewrite theories can be made easier by providing a sophisticated graphical tool called \narval\ that supports the   fine-grained  inspection  of Maude symbolic computations. This paper is under consideration for acceptance in TPLP.
\end{abstract}
\begin{keywords}
Symbolic reachability analysis, narrowing, equational unification, Maude, rewriting logic 
\end{keywords}

\vspace{-2ex}
\section{Introduction}\label{intro}
Maude \cite{maude07} is a high-performance implementation of rewriting logic, a simple extension of equational logic that models concurrent systems \cite{Meseguer12}. Maude seamlessly integrates: (i) functional, logic, concurrent, and object-oriented computations; (ii) rich type structures with sorts, subsorts, and operator overloading; and (iii) equational reasoning modulo axioms such as associativity (A), commutativity (C), and unity (U) of functions. With regard to the language performance, Maude was ranked second (after Haskell) as the best performance language in a recent benchmarking of well-known algebraic, functional, and object-oriented languages\footnote{CafeOBJ, Clean, Haskell, LNT, LOTOS, Maude, mCRL2, OCaml, Opal, Rascal, Scala, SML, Stratego / XT, and Tom; see references in \cite{GTA18}.} carried out in \cite{GTA18}. Because of its efficient rewriting engine and its meta-language capabilities, Maude is an excellent instrument for creating rich executable environments for various logics, programming languages, and tools (e.g.,\ \cite{ABFS16-tplp}). 

A rewrite theory (or Maude program) is a triple $\cR = (\Sigma, E, R)$ where $\Sigma$ is a signature that contains the program operators together with their type definition, $E$ is a collection of (possibly conditional) $\Sigma$-equations (so that $(\Sigma,E)$ is an equational theory) that models system states as terms of an algebraic data type (initial algebra), ${\cal T}_{\Symbols/E}$, and $R$ is a set of (possibly conditional) rewrite rules that define concurrent transitions between states. The equational theory $E$ is generally decomposed as a disjoint union $E=E_{0} \uplus Ax$, where the set $E_{0}$ consists of (conditional) equations and membership axioms (i.e.,\ axioms that assert the type or {\it sort} of some terms) that are implicitly oriented from left to right as rewrite rules (and operationally used as simplification rules), and $Ax$ is a set of commonly occurring algebraic axioms such as associativity, commutativity, and identity that are implicitly expressed as function attributes and are mainly used for $Ax$-matching (e.g.,\ assuming a commutative binary operator $*$, the term $s(0)*0$ matches within the term $X*s(Y)$ \emph{modulo} the commutativity of symbol $*$ with matching substitution $\{X/0,Y/0\}$).

The rewrite theory $\cR$ specifies a concurrent system that evolves by rewriting states using {\it equational rewriting}, i.e.,\ rewriting with the rewrite rules in $R$ {\it modulo} the equations and axioms in $E$ \cite{Meseguer92}. For instance, consider the sort {\textit{Int} for integer numbers that are generated from the constant $0$, the successor operator {\em s}, and the predecessor operator {\em p}, and that are endowed with the commutative addition operator $+$. Consider the (partial) specification of integer numbers defined by the equations $E_{0}= \{ (e1)~ X+0 = X, (e2)~ X+s(Y) = s(X+Y), (e3)~ p(s(X)) = X, (e4)~ s(p(X)) = X\}$, where variables $X$, $Y$ are of sort \textit{Int}, and $Ax$ contains the commutativity axiom $X+Y=Y+X$. Consider also a binary state constructor operator $\langle\_,\_\rangle: Int\ Int \rightarrow \mathit{State}$ for a new sort {\em State} that models a system of processes waiting to enter a critical section (in the first argument) and inside the critical section (in the second argument). The system state $t =\langle s(0),s(0) + p(0)\rangle$ can be rewritten with the following rule (denoting that a waiting process is entering the critical section)
\begin{align}
\langle s(X),Y\rangle \Rightarrow \langle p(s(X)),s(Y)\rangle\tag{$r1$}\label{eq:rule1}
\end{align} 

\noindent
which yields the state $\langle 0,s(0)\rangle$. This is essentially done in Maude by first simplifying the input state $t$ (with the equations $E_0$ modulo $Ax$) to its irreducible form\footnote{Note that the term $t =\langle s(0),s(0) + p(0)\rangle$ is first simplified into $\langle s(0),s(p(0)+0))\rangle$ by reducing the subterm $s(0)+p(0)$ of $t$ with the (implicitly oriented) equation $(e2)$ in $E_{0}$ \emph{modulo} the commutativity of $+$. The term $\langle s(0),s(p(0)+0))\rangle$ is then further simplified into $\langle s(0),s(p(0))\rangle$ by reducing the subterm $s(p(0)+0)$ with $(e1)$. And then $\langle s(0),s(p(0))\rangle$ is simplified into the irreducible term $t_{\downarrow_{E_0,Ax}}=\langle s(0),0\rangle$ by reducing the subterm $s(p(0))$ with $(e4)$.} $t_{\downarrow_{E_0,Ax}}=\langle s(0),0\rangle$. Then $t_{\downarrow_{E_0,Ax}}$ is rewritten into $t'=\langle p(s(0)),s(0)\rangle$ by applying the rewrite rule \eqref{eq:rule1}. And, finally, $t'$ is also normalized with the equations $E_0$ (modulo $Ax$) to its irreducible form $t'_{\downarrow_{E_0,Ax}}=\langle 0,s(0)\rangle$ by applying equation $(e3)$ in $E_{0}$ to the subterm $p(s(0))$ of $t'$. In symbols, a rewrite step $t\stackrel{r}{\longrightarrow}_{\cR} s$ consists of the rewrite sequence $t\rightarrow_{E_0,Ax}^*(t_{\downarrow_{E_0,Ax}})\: \rightarrow_{\{r\},Ax} t'\rightarrow_{E_0,Ax}^*(t'_{\downarrow_{E_0,Ax}})$, with $t'_{\downarrow_{E_0,Ax}}=s$, and denotes a transition (modulo $E$) from state $t$ to state $s$ via the rewrite rule $r$ of $R$.

System computations (also called execution traces) correspond to equational rewrite sequences $t_0\stackrel{r_0}{\longrightarrow}_{\cR} t_1 \stackrel{ r_1}{\longrightarrow}_{\cR}\ldots$, where each $t_{i} \stackrel{r_i}{\longrightarrow}_{\cR} t_{i+1}$ denotes a transition (modulo $E$) from state $t_{i}$ to $t_{i+1}$ via the rewrite rule $r_i$ of $R$. The transition space of all computations in $\cR$ from the initial state $t_0$ can be represented as a {\em computation tree} whose branches specify all the system computations in $\cR$ that originate from $t_0$.

Rewriting and equational rewriting are of course symbolic reasoning methods, but Maude supports symbolic reasoning in a stronger sense by means of {\em (equational) unification and narrowing in the rewrite theory $\cR=(\Sigma, E, R)$}. Narrowing is a generalization of term rewriting that allows free variables in terms (as in logic programming) and handles them by using unification (instead of pattern matching) to non-deterministically reduce these terms. Originally introduced in the context of theorem proving, narrowing is complete in the sense of logic programming (computation of answers) and functional programming (computation of irreducible forms) so that efficient versions of narrowing have been adopted as the operational principle of so-called multi-paradigm (constraint, functional, and logic) programming languages (see, e.g., \cite{Hanus13}). In the last few years, there has been a resurgence of narrowing in many application areas such as equational unification, state space exploration, protocol analysis, termination analysis, theorem proving, deductive verification, model transformation, testing, constraint solving, and model checking of infinite-state systems. To a large extent, the growing interest in narrowing is motivated by the recent takeoff of symbolic execution applications and the availability of efficient narrowing implementations. Narrowing-based, symbolic reasoning methods and applications in rewriting logic and Maude are discussed in \cite{Meseguer18-wollic}.

Similarly to equational rewriting, where matching modulo $E$ (or $E$-matching) is used, {\em equational} unification (or $E$-unification) is adopted (instead of standard, syntactic unification) in $(R,E)$-narrowing (i.e.,\ narrowing with the rules in $R$ modulo the equations and axioms in $E$). More precisely, $(R,E)$-narrowing in a rewrite theory $\cR=(\Sigma, E, R)$, with $E=E_{0} \uplus Ax$, is supported in Maude by means of a {\em three-level} machinery \cite{maude-manual}.

\begin{enumerate}
\item An $(R,E)$-narrowing step from $t_{1}$ to $t_{2}$ with a rule $l \Rightarrow r$ in $R$ is carried out by first performing $E$-unification between a subterm $s$ of the normalized version $t'_{1}$ of $t_{1}$, i.e., $t'_{1}= {t_{1}}_{\downarrow_{E_0,Ax}}$, and the left-hand side $l$ of the rule. The term $t_2$ is obtained from $t'_1$ by first replacing $s$ in $t'_1$ with the right-hand side $r$, then instantiating the yielded term with the computed $E$-unifier, and finally normalizing the resulting term with $E_{0}$ modulo $Ax$. Note that the rule may have extra variables in its right-hand side.
\item In turn, each $E$-unification problem $s=^{?}_{E} l$ of Point 1 is solved by using {\em folding variant} narrowing (in short, FV-narrowing) in the equational theory $(\Sigma,E)$, which is an equational narrowing strategy that computes a finite and complete set of $E$-unifiers for $s=^{?}_{E} l$ under suitable requirements \cite{ESM12}. The idea of FV-narrowing is to {\em equationally} narrow the term $s\: =\!?\!\!=\: l$ (that encodes the unification problem $s=^{?}_{E} l$) into an extra constant $tt$ in the rewrite theory $\cR_0 =(\Sigma\cup\{=\!?\!\!=,tt\},Ax,\vec{E_{0}}\cup\{\epsilon\})$, where $\vec{E_{0}}$ results from explicitly orienting the equations of $E_{0}$ as rewrite rules. Following \cite{MH92}, the extra rewrite rule $\epsilon=(X\: =\!?\!\!=\: X \Rightarrow tt)$ is added\footnote{Actually, in an order-sorted setting, multiple equations  are used to cover any possible sort in $\cR$. See Section \ref{sec:imple} for a detailed discussion.} to $\vec{E_{0}}$ in order to mimic unification of two terms (modulo $Ax$) as a narrowing step that uses $\epsilon$. For example, by using $\epsilon$, the term $s(0)*0\:=\!?\!=\:U*s(V)$ FV-narrows to {\tt tt} (modulo commutativity of $*$), and the computed narrowing substitution does coincide with the unifier modulo C of the two argument terms, i.e.,\ $\{U/0,V/0\}$.
\item For each folding variant narrowing step using a rule in $\vec{E_0}$ modulo $Ax$ in Point 2, $Ax$-unification algorithms are employed, allowing any combination of symbols having any combination of associativity, commutativity, and identity \cite{DEEM+18}. 
\end{enumerate}

$(R,E)$-narrowing computations are natively supported\footnote{Maude 2.8 is currently available as Maude alpha version 121 for developers, and it will be officially released soon.} by Maude version 2.8 for unconditional rewrite theories. Following the previous example, the input state $\langle 0,0+s(Z)\rangle$ $(R,E)$-narrows to $\langle p(0),s(s(Z))\rangle$ with substitution $\{X/p(0),Y/s(Z)\}$ (first level), i.e.,\ an $E$-unifier of the normalized term $\langle 0,s(Z)\rangle$ and the left-hand side $\langle s(X),Y\rangle$ of the rewrite rule \eqref{eq:rule1} in the theory $\cR$ of the example above. More precisely, the $E$-unifier $\{X/p(0),Y/s(Z)\}$ is the computed narrowing substitution obtained by FV-narrowing the term $\langle 0,s(Z)\rangle$ $=\!?\!\!= \langle s(X),Y\rangle$ to $tt$. This is done by first narrowing the subterm $s(X)$ via the (renamed apart) program equation $(e4)~ s(p(X'))=X'$, with partially computed substitution $\theta_1=\{X/p(X')\}$, and then narrowing the resulting goal $\langle 0,s(Z)\rangle$ $=\!?\!\!= \langle X',Y\rangle$ by using equation $\epsilon$, with partially computed substitution $\theta_2=\{X'/0,Y/s(Z)\}$, which finally yields the $E$-unifier $\theta_1\theta_2=\{X/p(0),Y/s(Z)\}$ (second level). Note that purely syntactic unifiers $\theta_{1}$ and $\theta_{2}$ are computed by Maude's built-in $Ax$-unification in this FV-narrowing derivation since operators $0$, $s$ and $p$ obey no algebraic axioms (third level).

Note that the narrowing step from $\langle 0,0+s(Z)\rangle$ to $\langle p(0),s(s(Z))\rangle$ signals a possible programming error in rule \eqref{eq:rule1} since it shows that multiple processes might enter a critical section, simultaneously, and the number of processes in the waiting list is negative.

Analogously to rewriting, the search space of $(R,E)$-narrowing computations in $(\Sigma,E,R)$ (respectively, FV-narrowing computations in $(\Sigma,E)$) can be represented as a tree-like structure that we call $(R,E)$-narrowing (respectively, FV-narrowing) tree. When it is clear from the context, we simply write narrowing instead of $(R,E)$-narrowing or FV-narrowing. 

Maude (symbolic) computations are complex, textually large artifacts that are difficult to inspect. This paper describes \narval\ (\textit{Narrowing variant-based tool)}, a visual system for exploring all three levels of symbolic computations in Maude programs. This contribution is important because Maude lacks any graphical symbolic tracing facility that can help the user to advance stepwise through a given symbolic execution. This includes not only the inspection of partially computed substitutions, but also the internals of $Ax$-matching and equational simplification sequences as well as $Ax$-unification and equational unification sequences that are concealed within rewriting and narrowing algorithms and are hidden within Maude's symbolic execution machinery. In order not to jeopardize the language performance, many of the state transformations (using $E$) described above are internal and are never recorded explicitly in the symbolic trace; hence, any erroneous intermediate result is difficult to debug. Furthermore, Maude narrowing traces are either directly displayed or written to a file (in both cases, in plain text format) and are thus only amenable for manual inspection by the user. This is in contrast with the enriched views provided by \narval,\ which are complete (every single transition is recorded by default) and can be either graphically displayed or delivered in its meta-level representation, which is very useful for further automated manipulation. Also, the displayed view can be abstracted when deemed appropriate to avoid cluttering the display with unneeded details. Finally, important insights regarding the programs/theories can be gained from controlling the narrowing space exploration.

\narval\ complements two existing tools in the Maude ecosystem, namely, ANIMA \cite{ABFS15-jsc} and GLINTS \cite{ACES17-tplp}. ANIMA is a forward trace slicer and program animator for Maude that allows (ground) rewrite computations to be interactively simplified and explored. GLINTS is a graphical environment for interactive variant generation in an equational theory $(\Sigma, E)$ that can also be used to analyze whether a given theory satisfies the finite variant property, which is a fundamental requirement for the termination of FV-narrowing (and hence termination of equational unification). It is worth noting that neither ANIMA nor GLINTS support the interactive inspection of (three-level) narrowing computations in a rewrite theory $\cR=(\Sigma, E, R)$ and do not provide any $(R,E)$-narrowing symbolic reasoning functionality.

This paper summarizes our experience as follows: i) identifying what to visualize in terms of narrowing computations and how to represent each element; ii) showing how visualization can enhance program analysis and debugging; and iii) implementing the components of \narval.\ The rest of the paper is structured as follows. Section \ref{sec:prelim} introduces a leading example that will be used throughout the paper for describing the narrowing-based, symbolic reasoning capabilities of \narval.\ In Section \ref{sec:analysis}, we explain the core functionality of \narval\ that supports both symbolic search space exploration and interactive reachability analysis for Maude programs. We also show how such features can be used to diagnose and correct the Maude programs as well as to infer new formal descriptions that satisfy the user's intent. Section \ref{sec:extra} describes some additional tool features that allow the user to glimpse into the technical details of narrowing computations such as the fine-grained inspection of narrowing steps, the computation of equational unifiers, the exploration of different representations of Maude's narrowing and rewriting search spaces, and interactive visualization with source code inspection. In Section \ref{sec:imple}, we provide a description of the tool architecture and we overview the main implementation choices. Finally, some related work and further applications are briefly discussed in Section~\ref{sec:conc}.

\section{Software Systems as Maude programs: a Generic Grammar Interpreter}\label{sec:prelim}
Nondeterministic as well as concurrent software systems can be formalized through Maude {\em system modules} whose syntax is {\tt mod <NAME> is <SPEC> endm}, where {\tt <NAME>} is the module name and {\tt <SPEC>} encodes a given rewrite theory $\cR=(\Sigma,E_0\uplus Ax,R)$. Maude's syntax is almost self-explanatory, and here we just highlight the most relevant keywords that are used to specify $\Sigma$, $E_0$, $Ax$, and $R$ (for further details, please refer to \cite{maude-manual}). Sorts and operators of the signature $\Sigma$ are respectively declared by means of the keywords {\tt sort(s)} and {\tt op(s)}. Both prefix and mixfix notation can be used to specify operators: in the latter case, the special wildcard \verb+_+ is used as argument placeholder. Algebraic axioms in $Ax$ are attached to operator declarations via attributes: operator attributes {\tt assoc}, {\tt comm}, and {\tt id} respectively stand for associativity, commutativity, and identity. Subtyping relations are introduced by means of the {\tt subsort} keyword. Equations in $E$ are denoted by the {\tt eq} keyword. Similarly, keyword {\tt rl} defines rewrite rules in $R$, and the rule attribute {\tt narrowing} characterizes all and only those rewrite rules that can be used to perform $(R,E)$-narrowing steps (while the rest of rules in $R$ are only used for rewriting).

\begin{figure}[h!]
{\footnotesize
\begin{verbatim}
 1 mod GRAMMAR-INT is
 2   sorts Symbol NSymbol TSymbol String Production Grammar Conf .
 3   subsorts TSymbol NSymbol < Symbol < String .
 4   subsort Production < Grammar .
 5   ops 0 1 2 eps : -> TSymbol .
 6   ops S A B : -> NSymbol .
 7   op _@_ : String Grammar -> Conf .
 8   op _->_ : String String -> Production .
 9   op __ : String String -> String [assoc id: eps] .
10   op mt : -> Grammar .
11   op _;_ : Grammar Grammar -> Grammar [assoc comm id: mt] .
12   vars L1 L2 U V : String .
13   var G : Grammar . 
14   rl [apply] : ( L1 U L2 @ (U -> V) ; G) => ( L1 V L2 @ (U -> V) ; G) [narrowing] .
15 endm
\end{verbatim}
}
\caption{The {\tt GRAMMAR-INT} Maude module}\label{fig:grammar}
\end{figure}

To illustrate the novel features of the \narval\ tool, we consider, as a leading example, the Maude module {\tt GRAMMAR-INT} of Figure \ref{fig:grammar}, which encodes a concise, generic grammar interpreter from \cite{DEEM+18}. Interpreter states are specified by (possibly non-ground) terms of the form {\tt T\ @\ G}, where {\tt G} is an input grammar and {\tt T} represents an input string of terminal and nonterminal symbols. For the sake of simplicity, we provide three non-terminal symbols of sort {\tt NSymbol}: {\tt A}, {\tt B}, and {\tt S} (the start symbol), and four terminal symbols of sort {\tt TSymbol}: {\tt 0}, {\tt 1}, {\tt 2}, and the finalizing mark {\tt eps} (i.e., the empty string). We also declare sort {\tt Symbol} (that includes both {\tt NSymbol} and {\tt TSymbol}) as a subsort of {\tt String} so that strings can be simply built by using the (associative) juxtaposition operator \verb+__+ which also has {\tt eps} as unity element.

The input grammar {\tt G} is defined by means of the associative and commutative operator \verb+_;_+ with identity {\tt mt} (the empty grammar), which allows {\tt G} to be concisely specified as a multiset of productions $\tt U_1\rightarrow V_1 ; \ldots ; U_n\rightarrow V_n$, where each $\tt U_i\rightarrow V_i$ denotes a production rule of {\tt G}. Note that $\tt U_i, V_i$ in each production $\tt U_i\rightarrow V_i$ can be any arbitrary string of symbols; thus, any kind of grammar of Chomsky's hierarchy can be formalized within our very compact interpreter ---from the simple Type-3 grammars that generate regular languages to the unrestricted Type-0 grammars that denote recursively enumerable languages.

The interpreter behavior is specified by a single rewrite rule that implements state transitions (namely, the rule identified by label {\tt apply} in line 14). More specifically, given a state {\tt W\ @\ G}, this rule non-deterministically applies the production rules of {\tt G} to the string of symbols {\tt W}, thus yielding a new state; this is done by rewriting any substring {\tt U} of {\tt W} by using the production {\tt U -> V} of {\tt G} (with {\tt W} being non-deterministically decomposed as {\tt L1 U L2} thanks to matching modulo associativity in strings, and the production {\tt U -> V} being automatically identified thanks to ACU-matching in the multiset that represents the grammar {\tt G}), and then proceeding with {\tt L1~V~L2}. Generating a string $st$ that belongs to the language of {\tt G} consists of rewriting the initial state {\tt S\ @\ G} until the final state {\tt $st$\ @\ G} is reached. Moreover, the very same rule can be also used to recognize that a given string belongs to the language of {\tt G}. Parsing a word {\tt w} according to{ \tt G} can be obviously defined by doing rewriting in the opposite direction, e.g.,\ by defining a new rule
	
\begin{verbatim}
  rl [parsing] : (L1 V L2 @ (U -> V) ; G) => (L1 U L2 @ (U -> V) ; G) 
\end{verbatim}
	
However, there is no need to introduce this rule in Maude since recognizing {\tt w} can be simply achieved by solving the {\em reachability goal} {\tt S\ @\ G } $\longrightarrow^*$ {\tt w\ @\ G}.
	
\begin{example}\label{ex:grammar} Consider the following Type-2 grammar {\tt G}
\begin{align*}\tt
 S \rightarrow \tt \  0\, S\, 1\ | \tt\ 1\, 0
\end{align*}
that generates the language $\tt \{0^n101^n\ |\ n\geq 0\}$. Then, the language string {\tt 001011} can be generated by the following rewriting sequence in module {\tt GRAMMAR-INT}:
\begin{quote}
{\tt S\ @\ G} $\:\longrightarrow\:$ {\tt 0S1\ @\ G} $\:\longrightarrow\:$ {\tt 00S11\ @\ G} $\:\longrightarrow\:$ 
{\tt 001011\ @\ G}
\end{quote} 
Also, solving the reachability goal {\tt S\ @\ G } $\longrightarrow^*$ {\tt 001011\ @\ G} proves that the string is grammatically correct, and moreover, solving {\tt S\ @\ G } $\longrightarrow^*$ {\tt 001W\ @\ G} binds variable {\tt W} with the string value $\tt 011$. Note that the form of reasoning given by (classical) reachability goals $t \longrightarrow^* \exists X\: t'$, with $X$ being the set of variables appearing in $t'$, does not involve any unification as no variables in the input term $t$ get instantiated (but only $E$-matching of subsequently rewritten forms of $t$ within the term $t'$ is performed); that is, no {\em narrowing} on $t$ is performed. This is in contrast with the {\em symbolic reachability analysis} based on narrowing that we describe in Section~\ref{sec:analysis}.
\end{example}

\section{Narrowing-based Symbolic Reachability Analysis with \narval\ }\label{sec:analysis}

Given a Maude program that encodes a rewrite theory $\cR=(\Sigma,E_{0} \cup Ax,R)$, and a (possibly) non-ground term $t$, the search space of $\cR$ that originates from $t$ can be symbolically explored by using $(R,E)$-narrowing. Indeed, $t$ represents an abstract characterization of the (possibly) infinite set of all of the concurrent states $[\![t]\!]$ (i.e.,\, all the ground substitution instances of $t$, or, more precisely, the $E_{0} \cup Ax$-equivalence classes associated to such ground instances) within $\cR$. In this scenario, each $(R,E)$-narrowing computation $\cC$ subsumes all of the rewrite computations that are ``instances'' of $\cC$ modulo $E_{0} \cup Ax$ \cite{maude-manual}. Therefore, the narrowing tree that stems from $t$ offers a compact, symbolic representation of the program behaviors for the different instances of $[\![t]\!]$.

More importantly, an exhaustive exploration of the $(R,E)$-narrowing tree of $t$ allows one to prove existential reachability properties of the form
\begin{equation}\label{eq:reach}
\exists X\; t \longrightarrow^{*} t' 
\end{equation}
with $t$ and $t'$ being two (possibly) non-ground terms ---called the {\em input} term and {\em target} term, respectively--- and $X$ being the set of variables appearing in both $t$ and $t'$. Roughly speaking, proving the logic formula (\ref{eq:reach}) amounts to determining whether there exists a state in the set $[\![t]\!]$ of instances of $t$ from which we can \emph{reach} a state in the set $[\![t']\!]$ of instances of $t'$ after a finite (possibly zero) number of narrowing steps with the rules of $R$ modulo the equational theory $E_0\uplus Ax$. Solving this problem means searching for a symbolic solution to it within $\cR$'s narrowing tree that originates from $t$ in a hopefully \emph{complete} way (so that, for any existing solution, a more general answer modulo $E_0\uplus Ax$ will be found).

Completeness of $(R,E)$-narrowing holds for topmost rewrite theories (i.e.,\ all rewrites occur at the term root). This class of theories is of primary importance in the Rewriting Logic framework since it has many practical applications (e.g., the analysis of security protocols; see \cite{Meseguer18-wollic}). More complex theories (e.g., topmost modulo $Ax$ theories, and Russian doll theories) can be easily transformed into equivalent, topmost rewrite theories \cite{ABS19}. 

The Maude 2.8 distribution provides the built-in \texttt{vu-narrow} command that allows the symbolic search space of a given term $t$ to be explored as well as sophisticated reachability properties to be investigated by incrementally visiting, in a breadth-first manner, the narrowing tree for $t$. Since the narrowing search may either never terminate and/or find an infinite number of solutions, two \emph{bounds} can be added to the \texttt{vu-narrow} command: a bound for the number of solutions requested, and another bound for the number of narrowing steps from the initial input term $t$ (i.e., a threshold depth on the $(R,E)$-narrowing tree is set to make the search finite). Unfortunately, \texttt{vu-narrow} outputs are given in a raw, often giant, text format that can be difficult to inspect and understand for non-experienced users. 

The \narval\ system described in this paper gracefully overcomes this drawback by providing a rich, graphical environment, where narrowing trees can be exhaustively and stepwisely explored by means of an intuitive point-and-click strategy, and solutions for reachability problems are automatically highlighted when progressively hit during the incremental construction of the narrowing tree. This exploration process is completely interactive, since the tree expansion is guided by the user who is free to select the tree nodes to be expanded without following a predefined search strategy. This is in contrast to Maude, which solves reachability goals by applying a breadth-first search strategy, where the tree level $n+1$ is visited only when level $n$ has been exhaustively explored. The interactive search keeps the size of the explored tree fragment small since only nodes of some value for the user are explored while uninteresting nodes are ignored. Nevertheless, as a useful shortcut, \narval\ also provides automated depth-$k$ expansion of tree nodes, as discussed in Section \ref{sec:extra}. In the sequel, we illustrate the core symbolic analysis features of \narval\ by using the generic grammar interpreter of Section \ref{sec:prelim}.
\subsection{Interactive Search Space Exploration}
\narval\ offers an interactive exploration facility for the incremental construction and visualization of narrowing trees. By simply selecting a node (i.e., state) $t$ in the frontier of the (current) tree $\cT$, all of the $(R,E)$-narrowing steps from $t$ are automatically computed and added to $\cT$, thereby providing an incremental expansion in amplitude of the original tree fragment. This feature can be particularly convenient when debugging or analyzing Maude programs. Let us see an example.
\begin{example}\label{ex:gen} 
Consider again the {\tt GRAMMAR-INT} module of Figure \ref{fig:grammar}, together with the Type-1 grammar {\tt G} 
\begin{align*}\tt
 S \rightarrow &\tt \ 0\, A\, 2\ |\ 0\, 2 \\
 \tt 0\, A \rightarrow &\tt \ 0\, 0\, A\, 2\ |\ 0\, 2
\end{align*}
which fails to generate the following language $\tt \{0 2\}\cup\{0^n12^n\ |\ n>1\}$ since one of its productions contains a small bug. The bug can immediately be detected by feeding \narval\ with the input interpreter state $s_1$
\begin{verbatim} 
 N:NSymbol @ (S -> 0 A 2) ; (S -> 0 2) ; (0 A -> 0 0 A 2) ; (0 A -> 0 2)
\end{verbatim} 
where {\tt N} is a ``logic'' variable of sort {\tt NSymbol}, and generating the narrowing tree fragment $\cT$ of Figure \ref{fig:faulty}.
\begin{figure}[h!]
\centering
\includegraphics[width=\textwidth]{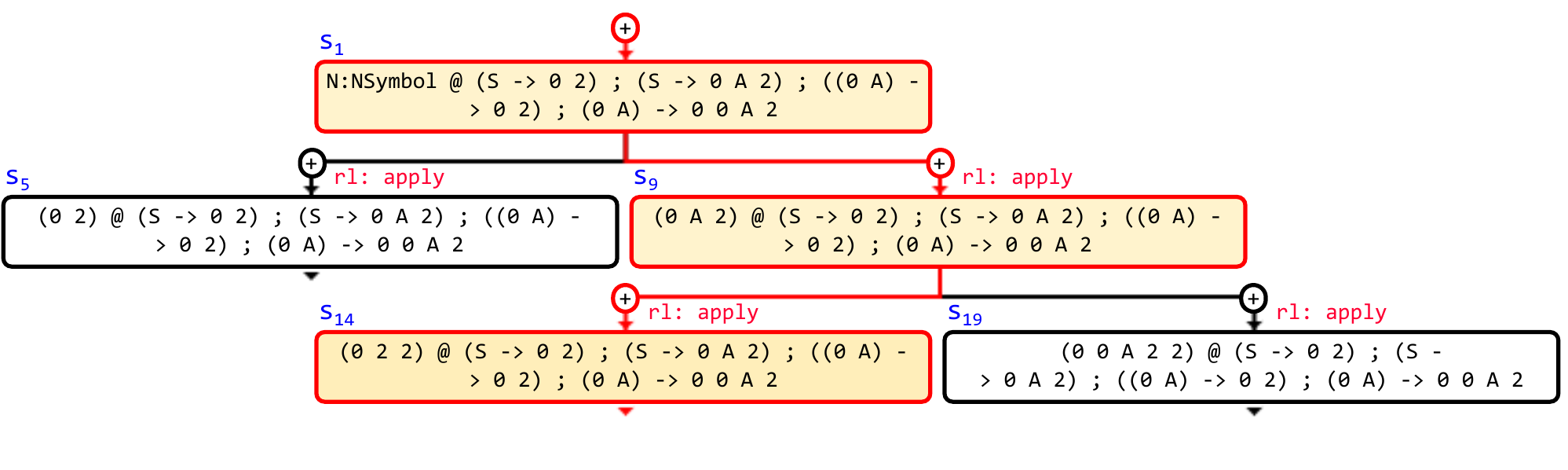}
\caption{Fragment of the narrowing tree computed by \narval\ in Example \ref{ex:gen}.}\label{fig:faulty}
\end{figure}
Indeed, each state {\tt w @ G} in $\cT$, with {\tt w} being a string of terminal symbols, indicates that {\tt w} is a word in the language accepted by {\tt G}, whenever the variable {\tt N} is bound to the non-terminal symbol {\tt S}. Now, observe that the $(R,E)$-narrowing step from state $s_{9}$ to state $s_{14}$ yields the undesired state {\tt 022 @ G} (with computed narrowing substitution {\tt \{N/S\}}) due to the application of the production ${\tt 0\, A \rightarrow \ 0\, 2}$, which replaces the nonterminal symbol {\tt A} with the erroneous terminal symbol {\tt 2} (which actually should be {\tt 1}). In this case, a fix can be obtained by replacing the faulty production with ${\tt 0\, A \rightarrow \ 0\, 1}$.
\end{example}

\subsection{Narrowing-based Reachability}\label{sec:symreach}
A reachability property $\exists X\: t \rightarrow^* t'$ is specified in \narval\ by simply providing the input and target terms, $t$ and $t'$, in the input phase. The property is then incrementally checked while expanding the $(R,E)$-narrowing tree of $t$ by simply $E$-unifying all of the nodes reached in the expanded tree fragment with the target term $t'$. Each branch in the tree that reaches a node $u$ that $E$-unifies with $t'$ is a narrowing computation that represents a constructive proof of the given property. Indeed, instantiating $t$ with the composition of the sequence of $E$-unifiers that enable each step in the narrowing computation from $t$ to $u$ (i.e.,\ the computed narrowing substitution $\sigma$), plus an $E$-unifier $\gamma$ for $u$ and $t'$, gives us a concrete rewrite sequence witnessing the existential reachability formula. To highlight the proof, $\narval$ automatically colors the node $u$ in green and the {\em reachability answer substitution} $\sigma\gamma$ is delivered.

As anticipated in Example \ref{ex:grammar}, the Maude module {\tt GRAMMAR-INT}, which was used as a pure, nondeterministic, word generator in Example \ref{ex:gen}, can also be employed as a parser that recognizes whether or not a word is in the language of a given grammar {\tt G}. Furthermore, more sophisticated reachability analyses than those shown in Example \ref{ex:grammar} can be achieved by using $(R,E)$-narrowing and its inherent logical program inversion capabilities, as shown in the following examples.
\begin{example}\label{ex:parsing}
Consider the following Type-2 grammar {\tt G}
\begin{align*}
\tt S \rightarrow\ &\tt 0\, S\, 0\ |\ 1\, S\, 1\ |\ eps
\end{align*} 
that generates the language of all even palindromes over the alphabet {\tt \{0,1\}}. Now, to show that the word {\tt 0110} is in the language of {\tt G}, we just need to feed \narval\ with the input term 
\begin{quote}
{\tt N:NSymbol @ (S -> 0 S 0) ; (S -> 1 S 1) ; (S -> eps) }
\end{quote}
and the target (ground) term {\tt\small 0 1 1 0 @ (S -> 0 S 0) ; (S -> 1 S 1) ; (S -> eps).} By exploring the $(R,E)$-narrowing tree, after a few tree expansions, we get the tree fragment of Figure \ref{fig:parser}
\begin{figure}[h!]
\centering
\includegraphics[width=\textwidth]{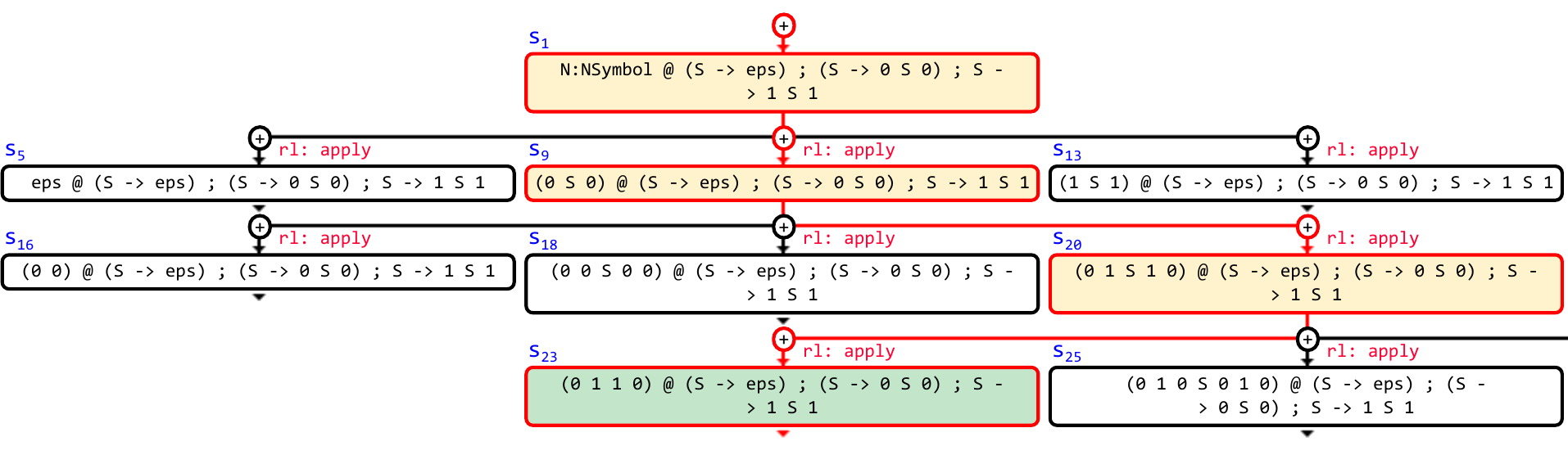}
\caption{Fragment of the narrowing tree computed by \narval\ in Example \ref{ex:parsing}.}\label{fig:parser}
\end{figure}
in which the green node $s_{23}$ shows that the word {\tt 0110} can be derived using {\tt G}. Moreover, by inspecting the details of the $(R,E)$-narrowing step from $s_{20}$ to $s_{23}$, we can discover that the reachability answer substitution includes the binding {\tt N:NSymbol/S}, which is correct and expected, since the word {\tt 0110} can only be generated starting from the grammar non-terminal symbol {\tt S}.
\end{example}

\begin{figure}[h!]
\centering
\includegraphics[width=\textwidth]{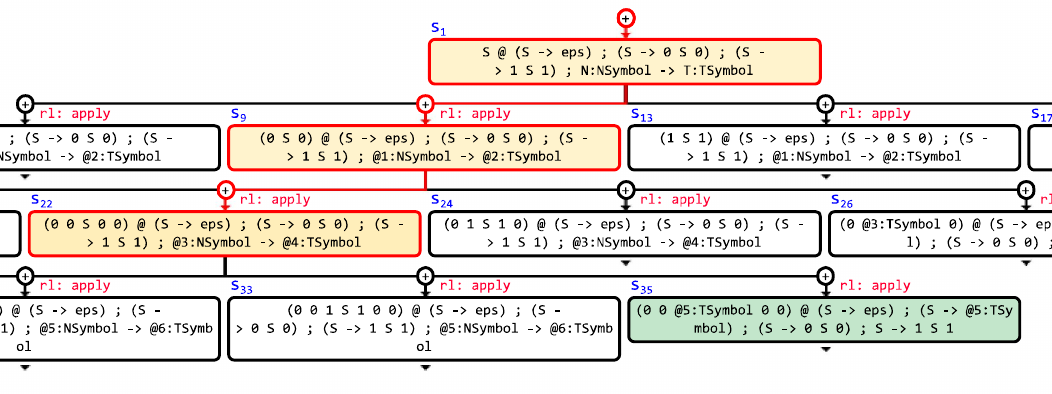}
\caption{Fragment of the narrowing tree computed by \narval\ in Example \ref{ex:missing}.}\label{fig:missingprod}
\end{figure}

Reachability symbolic analysis in \narval\ can also be a valuable tool for deriving new information from a given Maude program in order to automatically complete or mutate a given description w.r.t. the user's intent.
\begin{example}\label{ex:missing}
Consider the grammar {\tt G} of Example \ref{ex:parsing} and the palindrome {\tt 00100}, which does not belong to the language of {\tt G} since its length is odd. One may want to know what missing production is needed so that {\tt 00100} derives from the non-terminal symbol {\tt S}. This question can be automatically answered by feeding \narval\ with the input term
\begin{quote}{\tt S @ (N:NSymbol -> T:TSymbol) ; (S -> 0 S 0) ; (S -> 1 S 1) ; (S -> eps) 
}
\end{quote}
in which the original grammar specification has been augmented with a new, fully generic production {\tt N:NSymbol -> T:TSymbol}, and the target term 
\begin{quote}{\tt 0 0 1 0 0 @ (N:NSymbol -> T:TSymbol) ; (S -> 0 S 0) ; (S -> 1 S 1) ; (S -> eps)
}
\end{quote}
\end{example}

By using \narval,\ the user can interactively explore the symbolic search space and generate the tree fragment of Figure \ref{fig:missingprod} that includes the green node $s_{35}$ with the target expression. This node provides a solution for the considered reachability problem: indeed, the $E$-unification of the term in $s_{35}$ and the target term succeeds and computes the reachability answer substitution $\sigma=\tt \{N:NSymbol/S,T:TSymbol/1\}$ in Figure \ref{fig:reachsub}, which allows the missing production {\tt S -> 1} to be inferred from the instantiation with $\sigma$ of production {\tt N:NSymbol -> T:TSymbol}. 

\begin{figure}
\centering
\includegraphics[width=.9\textwidth]{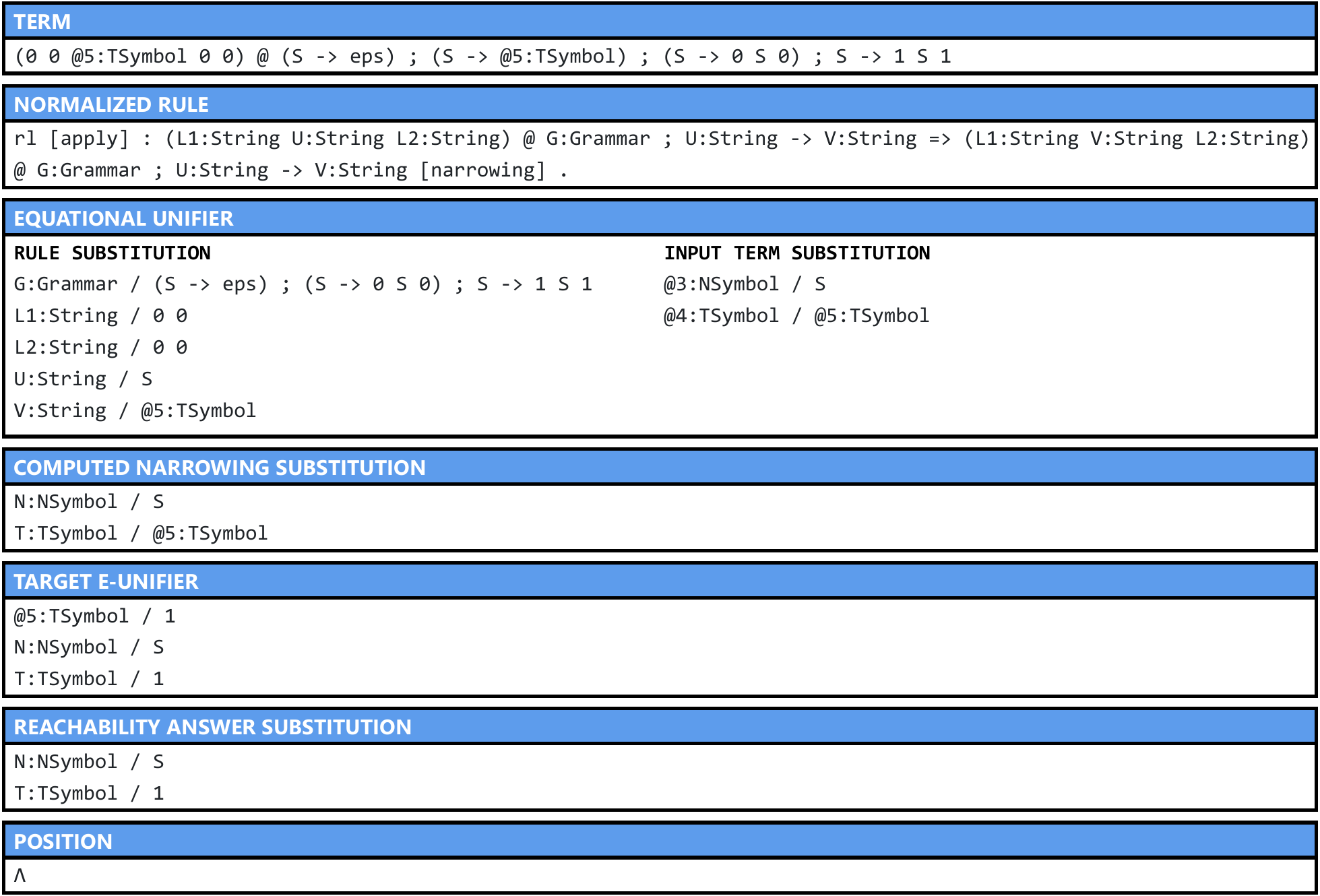}
\caption{Details of the narrowing step from $s_{22}$ to $s_{35}$ of Figure \ref{fig:missingprod}.}\label{fig:reachsub}
\end{figure}
\section{Additional Features of \narval}\label{sec:extra}
\paragraph{Execution modalities.} Besides the core, $(R,E)$-narrowing functionality that we discussed in Section \ref{sec:analysis}, which is available through the {\it Narrowing in a rewrite theory} execution mode, \narval\ supports three additional execution modalities: the {\em Rewriting} mode, the {\em FV-narrowing} mode, and the {\em Equational unification} mode. 

The rewriting mode allows the user to interactively explore the computation tree generated by Maude's {\em rewrite} engine that performs state transitions by non-deterministically rewriting system states modulo equations and axioms instead of using the much more involved $(R,E)$-narrowing relation. This option turns \narval\ into a program stepper that can be conveniently used to animate programs w.r.t. concrete inputs (i.e., ground input terms).

The FV-narrowing mode implements an inspection modality, based on the folding variant narrowing strategy of \cite{ESM12}, that only uses equations and axioms to narrow terms, thereby providing a means to analyze the purely equational search space of Maude programs. This modality serves also as a basis for the equational unification mode that inspects the insights of $E_0\uplus Ax$-unifiers that are computed by FV-narrowing and Maude's built-in $Ax$-unification algorithms. In fact, as explained in Section \ref{intro}, given a rewrite theory $\cR=(\Sigma,E,R)$, with $E=E_0\uplus Ax$, each $(R,E)$-narrowing step requires performing $(E_0\uplus Ax)$-unification between the term $t$ to be narrowed and the left-hand side $l$ of the applied rewrite rule. More precisely, this is done by executing a new \narval\ instance that explores the FV-narrowing tree rooted at $t\: =\!?\!=\: l$. Computed narrowing substitutions in successful tree branches (i.e., branches that end with the success constant $tt$) represent $(E_0\uplus Ax)$-unifiers for the unification problem $t=^?_{E_0\uplus Ax}l$. \narval\ automatically highlights, within the inspected FV-narrowing tree fragment, the tree branch that corresponds to the computation of the considered $(E_0\uplus Ax)$-unifier. Let us see an example.
\begin{example}\label{ex:unifier} 
Consider the $(R,E)$-narrowing step from state $s_{1}$ to state $s_{9}$ in the narrowing tree of Figure \ref{fig:parser}. The step uses the {\tt apply} rewrite rule of the {\tt GRAMMAR-INT} module and computes the equational unifier

{\small
\begin{verbatim}
{G:Grammar / (S -> eps) ; (S -> 1 S 1), L1:String / eps,
L2:String / eps, U:String / S, V:String / 0 S 0, N:NSymbol / S}.
\end{verbatim}
}

\noindent
To inspect the computation of this $E$-unifier, it suffices to right-click the state $s_{9}$ and select {\sf Inspect unifier} from the contextual menu. This action starts the exploration of the FV-narrowing tree that solves the equational unification problem between the state $s_{1}$ (that $(R,E)$-narrows into $s_{9}$) and the left-hand side of the {\tt apply} rule. Figure \ref{fig:unifier} shows a fragment of the FV-narrowing tree that includes the FV-narrowing computation of the equational unifier under examination (that is, the FV-narrowing computation from the root node $s_1$ to the green node $s_9$).
\begin{figure}[h!]
\centering
\includegraphics[width=\textwidth]{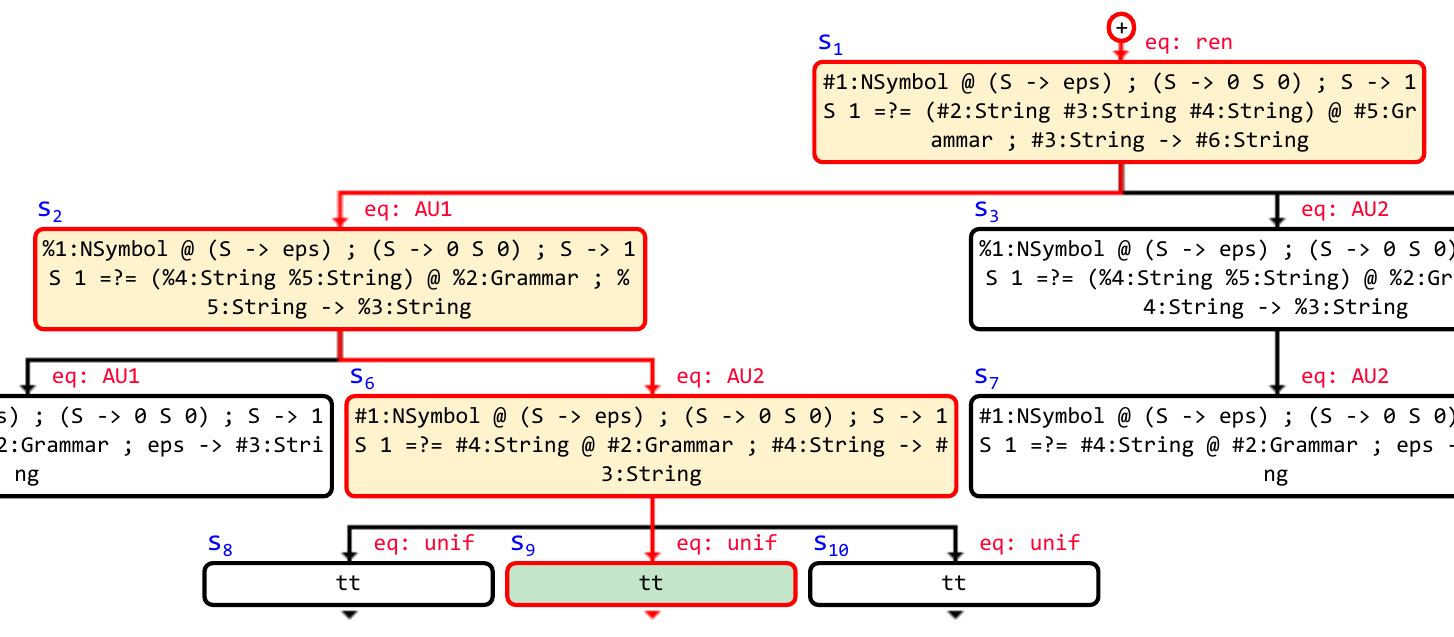}
\caption{Fragment of the FV-narrowing tree that computes the equational unifier in Example \ref{ex:unifier}.}\label{fig:unifier}
\vspace{-.5cm}
\end{figure}
\end{example}

\paragraph{Transition and computation information.}
By clicking any rule (edge) label in the narrowing tree, the user can obtain the complete information of the associated $(R,E)$-narrowing step. Specifically, the following information is provided: (i) the term $t$ that is yielded by the narrowing step; (ii) the (normalized version of the) rewrite rule applied; (iii) the position in $t$ where the narrowing step occurred ($\Lambda$ denotes the root position of $t$, while sequences of natural numbers are used to identify the positions of the proper subterms of $t$ in the usual way); (iv) the computed $(E_0\uplus Ax)$-unifier, which is organized in two disjoint substitutions: the {\em rule} substitution that includes the unification bindings that come from the variables in the rewrite rule, and the {\em input term} substitution that contains the bindings for the variables appearing in the input term to be narrowed; (v) the {\em computed narrowing} substitution, (vi) the target $E$-unifier (if any), that is, the $E$-unifier between $t$ and the target term of a reachability goal, and (vii) the reachability answer substitution (this field is only present if $t$ is a solution of a specified reachability property). For instance, Figure \ref{fig:reachsub} shows this transition information for the narrowing step from state $s_{22}$ to state $s_{35}$ of the narrowing tree of Figure \ref{fig:missingprod}. Similarly, the user can also access the details of FV-narrowing steps that are performed by means of dedicated $Ax$-unification algorithms.

\paragraph{Enriched views.}
\narval\ supports two distinct views of the $(R,E)$-narrowing and FV-narrowing trees: namely, the standard view and the instrumented view. The standard view (which is the default mode) focuses on the narrowing steps, whereas the instrumented view completes the picture with all of the internal reduction steps that are performed, using equations, axioms, and Maude built-in operations up to reaching the normalized form of each term. The instrumented view can be locally enabled on a selected narrowing step by pressing the {\sf +} symbol labelling the associated edge in the tree. For instance, Figure \ref{fig:instrumented} illustrates the instrumented view for the $(R,E)$-narrowing step from $s_1$ to $s_5$ in Figure \ref{fig:parser}. Observe that the instrumented sequence of light blue nodes in Figure 7 rewrites the term {\tt eps eps eps} into its normalized form {\tt eps} by using (an explicit equational representation of) the unity axiom for the operator \verb+__+ (see Section \ref{sec:imple} for further details).
\begin{figure}[h!]
\centering
\includegraphics[width=\textwidth]{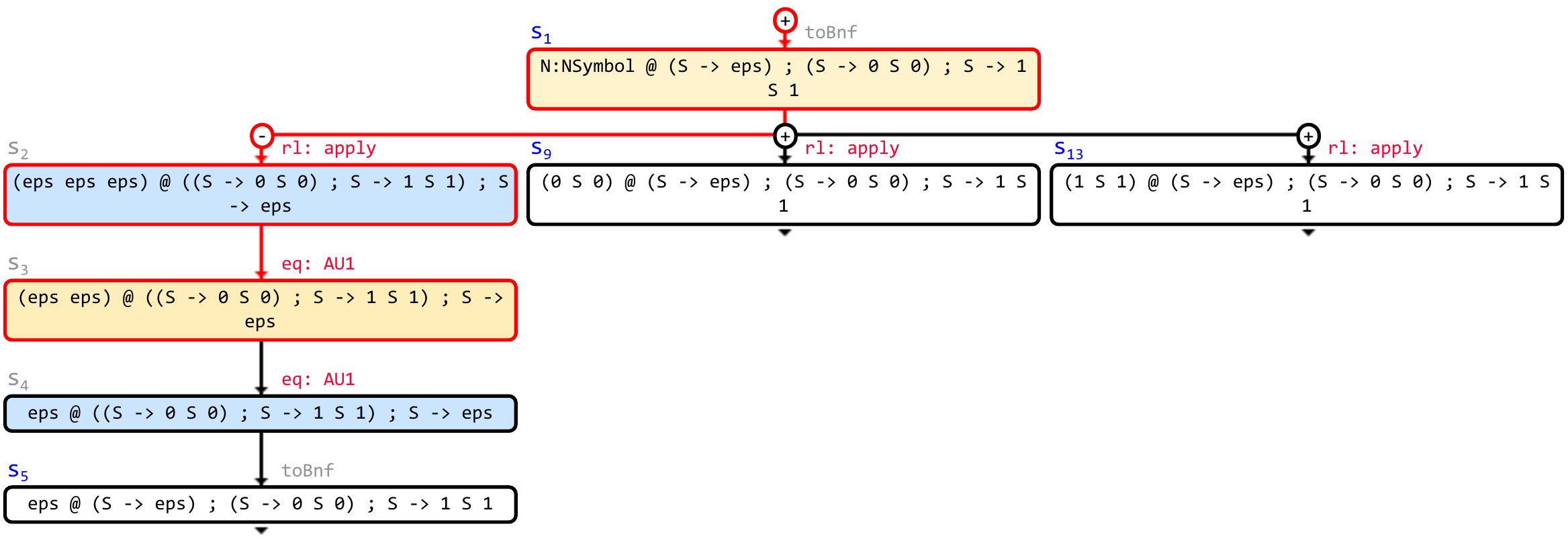}
\caption{Instrumented view for the $(R,E)$-narrowing step from $s_1$ to $s_5$ appearing in Figure \ref{fig:parser}.}\label{fig:instrumented}
\end{figure}

\paragraph{Additional navigation capabilities.}
\narval\ encompasses some additional features to improve the user experience while navigating (narrowing) trees. The user can automatically explore multiple levels of the tree by right-clicking with the mouse on a node $s$ and then selecting {\sf Expand Subtree} from the contextual menu. This option allows the user to automatically unfold, up to a given depth $k$, for $k \leq 5$ (with default depth $k = 3$), the subtree hanging from the considered node $s$ by following a breadth-first strategy. Dually, a subtree rooted at $s$ can be folded into $s$ by means of the {\sf Fold Node} option. 

Common actions like zooming in/out, dragging or moving the tree via arrow keys are supported. Also, when a tree node is selected, the position of the tree on the screen is automatically rearranged to keep the chosen node at the center of the scene. Finally, several data that decorate the narrowing tree (e.g. state/rule labels and substitution information) could be toggled on and off to focus the user attention on selected aspects of the explored search space.

Finally, even if the narrowing search space for a given input term is hierarchically organized as a tree in order to systematize its exploration, \narval\ additionally supports a graph-like representation where equivalent system states (i.e.,\ states that are equal modulo variable renaming and algebraic axioms) are grouped together in a single state representative. 

\section{Implementation}\label{sec:imple} 
\narval\ does not simply process Maude's output for the different symbolic features (rewriting, narrowing, and equational unification) to draw an exploration tree. Instead, it builds on top of Maude's reflective capabilities for reproducing in full detail every rewrite step, FV-narrowing step, and $(R,E)$-narrowing step in a rewrite theory $\cR$. There are many aspects to be dealt with symbolic computations, such as proper variable renaming, handling substitutions modulo axioms, or avoiding variable capturing (by providing different families of variable indices) that are made completely explicit in \narval, while Maude hides them inside its narrowing machinery. 

Furthermore, \narval\ performs a program transformation to support the in-depth analysis of equational unification. Equational unification in Maude is available by means of the {\tt variant unify} command and its corresponding meta-level operation {\tt metaVariantUnify} (see Chapter 12.9 of \cite{maude-manual}). However, these commands provide poor insight into the internal reasoning yielding their results. \narval\ automatically transforms the input program in order to explicitly encode the equational unification problem to be solved within the program itself and then uses Maude's standard functionality to carry out the inspection of the equational unification computation. Specifically, the input program is complemented by adding the binary, polymorphic operator \verb+=?=+ that is used to specify unification problems, and the constant operator {\tt tt} that represents success in computing an equational unifier. These operators are declared as follows:
{\small
\begin{verbatim}
    op _=?=_ : Universal Universal -> [Bool] [poly (1 2)] .
    op tt : -> [Bool] . 
\end{verbatim}
}

\noindent
where {\tt Universal} denotes a placeholder for any known sort, {\tt [Bool]} is the {\em kind} for the Boolean sort\footnote{A kind can be seen as an error supersort of a given sort.}, and {\tt poly (1 2)} specifies that both arguments of {\tt =?=} are polymorphic.

The input program is then augmented with a set of unification equations (one for each kind in the program) of the form: \verb+ eq [unif] : X =?= X = tt [variant]+ where {\tt X} is a variable of the appropriate kind. The equational attribute {\tt variant} distinguishes those equations in the equational theory that can be used for equational unification via FV-narrowing (while the rest of equations in the theory are only used for equational simplification).

Finally, an additional program transformation sketched in \cite{DEEM+18} is required when the input program includes AU operators, that is, operators equipped only with the {\tt assoc} and {\tt id} equational attributes. This is because the current version of Maude does not natively support AU-unification\footnote{More precisely, the latest version of Maude \cite{maude-manual} supports built-in, $Ax$-unification for the following combinations of equational attributes: the {\tt assoc} attribute (A), the {\tt comm} attribute (C), the {\tt assoc} {\tt comm} attributes (AC), the {\tt assoc} {\tt comm} {\tt id} attributes (ACU), the {\tt comm} {\tt id} attributes (CU), the {\tt id} attribute (U), the {\tt left id} attribute (Ul), and the {\tt right id} attribute (Ur).}. Nonetheless, unification modulo associativity and unity can still be performed by resorting to the following program transformation that replaces any operator declaration

{\small
\begin{verbatim} 
  op f : S S -> S [assoc id: idname] .
\end{verbatim} 
}

\noindent
with the following Maude code

{\small
 \begin{verbatim} 
  op f : S S -> S [assoc] .
  eq [AU1] : f(idname,X:S) = X:S [variant] .
  eq [AU2] : f(X:S,idname) = X:S [variant] .
  eq [AU3] : f(X:S,idname,Y:S) = f(X:S,Y:S) [variant] .
\end{verbatim}
}

\noindent
that declares the associative operator {\tt f} and provides an explicit equational definition for the identity element {\tt idname} that can be used to implement AU-unification via FV-narrowing and the Maude, built-in, A-unification algorithm. Note that the transformation includes the usual equations that are required to model left- and right-identity (namely, {\tt AU1} and {\tt AU2}), plus the additional equation {\tt AU3} that is used to enforce the coherence property between the rewrite rules and equations of the input program. Coherence is an essential executability requirement that guarantees the completeness of the Maude rewrite strategy $\longrightarrow_\cR$ that implements rewriting modulo equations and axioms (for further details, see Chapter 5 of \cite{maude-manual}).

\begin{example}\label{ex:transformation}
Consider the sort and subsort declarations of the Maude module {\tt GRAMMAR-INT} of Figure \ref{fig:grammar}, which identify four kinds, namely, \verb+[Bool]+, \verb+[String]+, \verb+[Grammar]+, and \verb+[Conf]+. 

\narval\ automatically augments the signature of {\tt GRAMMAR-INT} by adding the two abovementioned operator declarations for operators \verb+=?=+ and \verb+tt+, and extends the original equational theory with the following (variant) equations, one for each kind of the program:

{\small\begin{verbatim}
  eq X:[Bool] =?= X:[Bool] = tt [variant] .
  eq X:[String] =?= X:[String] = tt [variant] .
  eq X:[Grammar] =?= X:[Grammar] = tt [variant] .
  eq X:[Conf] =?= X:[Conf] = tt [variant] .
\end{verbatim}
}

\noindent
Moreover, it automatically replaces the AU operator \verb+__+ in line 9 with the following code that provides an explicit equational representation of the identity element {\tt eps}.

{\small \begin{verbatim} 
  op __ : String String -> String [assoc] .
  eq [AU1] : eps X:String = X:String [variant] .
  eq [AU2] : X:String eps = X:String [variant] .
  eq [AU3] : X:String eps Y:String = X:String Y:String [variant] .
\end{verbatim}
}

\noindent
\end{example}
The implementation of the proposed program transformations heavily relies on Maude reflective capabilities that correctly simulates the relevant metatheoretic features of Maude such as loading a module, evaluating a term, generating variants, computing unifiers, or performing narrowing steps. In our scenario, reflection is systematically used to handle programs as regular data structures through their meta-level representations that can be accessed and manipulated by using basic data operations such as list/multiset insertions and deletions. This way, a program can be easily transformed by first lifting it to its meta-level representation, and then adding, deleting or changing program items such as sort/operator declarations, equations and rewrite rules through data insertions and deletions. Furthermore, the resulting (meta-)program can be visualized in a friendly, source-level representation by means of the {\sf Show (transformed) program} option that efficiently implements (meta) string conversion in C++.

\paragraph{Architecture of \narval.}\label{sec:arch} 
The \narval\ tool has been implemented as a web application and is publicly available at \url{http://safe-tools.dsic.upv.es/narval}. \narval 's architecture consists of three main components (i.e., \narval 's core, client, and web services) that are implemented by combining a number of different technologies.

Firstly, the underlying rewriting and narrowing machinery of \narval 's core has been implemented in a custom version of Maude, named Mau-Dev \cite{maudev}, which provides extensions for some critical operators such as {\tt metaReducePath} (see \cite{ABFS15-jsc}) or {\tt metaGetVariant} (see \cite{ACES17-tplp}). These extensions are necessary to fully reproduce in detail the internal reasoning modulo axioms of Maude, which is only retrievable as raw text by using the interpreter's built-in (rewriting) debugger. \narval 's core consists of approximately 1800 lines of Maude and C++ source code. 

Secondly, the user-friendly graphical user interface of \narval's client has been implemented by using CSS, HTML5, and Javascript. Specifically, \narval 's graphical controls have been implemented by using the latest available version of Bootstrap 4, and the graph visualization feature is powered by the D3 for Data-Driven Documents library, which provides a representation-transparent approach to data visualization for the web. Without including these libraries, \narval's client consists of approximately 4400 lines of original HTML, CSS, and Javascript source code.

Finally, \narval's web service connects the core component of the system to the client user interface and consists of 10 RESTful Web Services that are implemented by using the JAX-RS API for developing Java RESTful Web Services (around 900 lines of Java source code).

\section{Conclusion and Related Work}\label{sec:conc}
Our main motivation for developing \narval\ was to assist users in analyzing complex software models described in Maude; however, the tool can also be used in training and education by showing the process and result of symbolic executions in a stepwise manner.

There are few tools in the literature for visualizing symbolic execution trees or narrowing trees. Symbolic execution \cite{King76} is a program analysis technique that is based on the interpretation of a program with symbolic values. Hahnle et al. implemented a tool, called visual symbolic state debugger, which can be used to debug sequential Java applications visually by using a symbolic execution tree \cite{HBBR10}. This makes it easy for the bug hunter to comprehend intermediate states and the actions performed on them in order to find the origin of a bug. SEViz \cite{HAZ15} is another tool for interactively visualizing symbolic executions as symbolic execution trees for simple .NET Framework programs. The visualization serves as a quick overview of the whole execution and helps to enhance the test input generation.

To our knowledge, \narval\ is the first graphical tool for the symbolic analysis of Maude rewrite theories. As future work, we plan to extend the trace slicing facility of \cite{ABFS15-jsc} to symbolic traces so that \narval\ can automatically produce reduced versions of narrowing computations according to a given slicing criterion, yielding more compact narrowing representations.

\vspace{-2ex}
\bibliographystyle{acmtrans}
\bibliography{biblio}
\label{lastpage}
\end{document}